# GRAZSCHUMPETERCENTRE

## *GSC Discussion Paper Series*

Paper No. 28

---

## Opinion Dynamics with Conflicting Interests

Patrick Mellacher[1]

[1] University of Graz, Graz Schumpeter Centre,

Universitätsstraße 15/FE, A-8010 Graz

### Abstract

I develop a rather simple agent-based model to capture a co-evolution of opinion formation, political decision making and economic outcomes. I use this model to study how societies form opinions if their members have opposing interests. Agents are connected in a social network and exchange opinions, but differ with regard to their interests and ability to gain information about them. I show that inequality in information and economic resources can have a drastic impact on aggregated opinion. In particular, my model illustrates how a tiny, but well-informed minority can influence group decisions to their favor. This effect is amplified if these agents are able to command more economic resources to advertise their views and if they can target their advertisements efficiently, as made possible by the rise of information technology. My results contribute to the understanding of pressing questions such as climate change denial and highlight the dangers that economic and information inequality can pose for democracies.

**JEL-Codes**: C63,D31,D62,D72,P16

**Opinion Dynamics with Conflicting Interests**

First version: 4[th] of November 2021

This version: 12[th] of November 2021

Patrick Mellacher[1]

[1] University of Graz

Graz Schumpeter Centre

patrick.mellacher@uni-graz.at

**Abstract**: I develop a rather simple agent-based model to capture a co-evolution of opinion formation, political decision making and economic outcomes. I use this model to study how societies form opinions if their members have opposing interests. Agents are connected in a social network and exchange opinions, but differ with regard to their interests and ability to gain information about them. I show that inequality in information and economic resources can have a drastic impact on aggregated opinion. In particular, my model illustrates how a tiny, but well-informed minority can influence group decisions to their favor. This effect is amplified if these agents are able to command more economic resources to advertise their views and if they can target their advertisements efficiently, as made possible by the rise of information technology. My results contribute to the understanding of pressing questions such as climate change denial and highlight the dangers that economic and information inequality can pose for democracies.

# 1 Introduction

Recent research shows that the oil industry has been informed of the impact of carbon dioxide on global warming since the 1950s (Franta 2018). Instead of taking early action to reduce its impact, however, this industry invested millions of dollars to spread misinformation about climate change (Frumhoff et al. 2015; Bonneuil et al. 2021).

Only recently, the tide has started to turn and action against climate change has begun to enjoy widespread popular support. One of the reasons for this change was that certain features of the climate crisis – such as the increase in global temperature – became virtually undeniable. Nevertheless, the "yellow vests" movement in France showcased the difficulties to enact particular types of environmental reforms. One crucial aspect of this example seems to be that a well-informed minority has been able to influence public opinion in a way that allowed it to maintain its high profits for a protracted period of time.

The general theme of a minority that is able to influence policy in a way that it harms the majority is the natural habitat of conspiracy theorists. However, it has also been recognized as an important theme by influential thinkers across the political spectrum. For instance, Buchanan and Tullock (1965) contended that special interest groups invest in "political organization" to influence governmental activity in order to advance their interests. Olson (1965) further claimed that it is more difficult to organize large groups due to a free-rider problem. Thus, smaller groups are able to advance their interests more effectively. On the other side of the ideological spectrum, Marx and Engels (e.g. 1969) argued that political power arises from the class structure of society, and the economic power of a small capitalist class thus manifests itself in and is reinforced by political power.

Nevertheless, formal studies of opinion dynamics, pioneered by French (1957), focus on the case of converging interests (Acemoglu et al. 2010) or are silent about interests, even if they account for the possibility of mis- and disinformation (e.g. Acemoglu and Ozdaglar 2011; Douven and Hegselmann 2021; Rajabi et al. 2020).[1] This silence prevails in spite of the fact that a significant fraction of this literature is devoted to studying the fragmentation and polarization of opinions, following the seminal contribution by Hegselmann and Krause (2002), and has made important advances to understand this phenomenon since (e.g. Acemoglu et al. 2010; Acemoglu et al. 2013; Banisch and Olbrich 2021; Feliciani et al. 2017; Pulick et al. 2016; Schweighofer et al. 2020; Siedlecki and Weron 2016).

In particular, this strand of the literature is able to explain the polarization of opinion in *absence* of a polarization of interests. Polarization/Fragmentation of opinions emerges in these models from at least one of the following assumptions: First, the ability of agents to influence each other is limited by a bounded confidence interval, i.e. agents whose opinions differ too drastically do not influence their respective opinions (Hegselmann and Krause 2002). Second, some agents are "stubborn" (Acemoglu et al. 2010), "campaigners" (Douven and Hegselmann 2021), "conspirators" or "inoculators" (Rajabi et al. 2020) and do not change their opinions. Third, social interactions are homophilic (Mäs and Flache 2013). Fourth, agents experience a reward when communicating their own views, which depends on the views held by their social neighborhood (Banisch and Olbrich 2021). Again, the social network needs to incorporate some form of homophily or disconnectedness in order to produce opinion polarization (ibid). Fifth, agents engage in "biased assimilation" i.e. process new information in a biased way that favors their initial beliefs (Dandekar et al. 2013). Sixth, agents may "push away" other agents from their own opinion (Flache and Macy 2011; Feliciani et al. 2017), either because those agents are ideologically too distant or because they exhibit anticonformity against outsiders (Siedlecki and Weron 2016). Finally, Schweitzer et al. (2020) show that a co-evolution of interpersonal attitudes and opinions can explain the polarization (and clustering) of opinions in a multi-dimensional opinion space.

---

[1] For instance, Acemoglu and Ozdaglar (2011) show that "forceful agents" who spread misinformation can cause a society to converge around a wrong belief. However, they do not argue *why* these agents may spread misinformation apart from "stubbornness".

In contrast to this literature, I seek to study how a *polarization of interests* may affect opinion formation. In particular, my contribution to this literature is the following:

I introduce conflicting interests into a simple opinion dynamics model. Interests matter in my model since a) agents may receive a signal from the outside world informing them about their true interests and b) because agents receive payoffs according to a vote conducted by the agents based on their beliefs. I further allow one type of agents to use their payoffs in order to advertise their views, and I finally investigate the impact of different advertisement strategies. Thus, my model offers a simple account of the interplay between opinion formation, political decision-making and economic processes.

I furthermore contribute to the literature on the causes of economic inequality using agent-based models, which currently focuses on either social networks (e.g. Gemkow and Neugart 2011: Dawid and Gemkow 2014) and/or technology as drivers of inequality (e.g. Carvalho and Di Guilmi 2020; Dawid et al. 2018; Dawid and Hepp 2021; Mellacher and Scheuer 2021; Mellacher 2021) by introducing a political source of inequality.

The rest of this paper is structured as follows: The next section describes the model in detail. The third section discusses its results. Section four concludes.

**2 Model**

The model is inhabited by two types of agents: $n_A$ agents of type A (that we can imagine, in the context of climate change, as polluters) and $n_B$ agents of type B (non-polluters). Agents of type A gain profits while causing external costs to agents of type B, but are constrained in doing so by laws that are democratically determined. Agents hold beliefs about their interests, which do not necessarily correspond to their actual interests. These beliefs may be influenced by a) discussions with other agents, b) advertisement/propaganda and c) information from the outside world. This model is discrete-time, implemented in NetLogo (Wilensky 1999) and the source code is available as open source.[2] During each time step of the simulation, the following sequence of events is computed:

1.) Agents try to convince each other (see 2.1)
2.) Agents consume advertisement (see 2.2)
3.) A fraction of agents receives information about their true interests (see 2.3)
4.) Agents vote on the law (see 2.4)
5.) Agents of type B obtain profits based on the current law (see 2.5)

While it is obvious to interpret this setup as an analysis of environmental damage, the model can also serve to illustrate the politico-economic dynamics of conflicting interests in a more general way. For instance, we could think of the relationship between a special interest group and the rest of the population, or the role that labor laws can play to limit exploitative work relations.

*2.1 Agent beliefs*

In order to simplify our problem, we can imagine the law governing the relation between the two types of agents $l_t$ as a continuum between 0 and 1, where 0 implies that agents of type A are not allowed to impose any external costs on other agents (e.g. pollution is not allowed at all), while 1 allows them to reap maximum profits. Each agent has a belief $b_{i,t}$ about the optimal law, which does not need to correspond with the law that is actually optimal for this agent.

---

[2] Publicly available after peer-review.

Agents are connected to each other via a social network, where they try to convince each other about their beliefs. More specifically, each agent $i$ influences the opinion of another agent $j$ to whom she or he is connected according to the following equation, where $b_{i,t}^*$ and $b_{j,t}^*$ are the respective current beliefs and $\mu$ is an influence parameter.

$$b_{j,t}^* = b_{j,t}^* + \mu(b_{i,t}^* - b_{j,t}^*) \tag{1}$$

The current belief $b_{i,t}^*$ is initialized as the belief held by agent i in the previous period $b_{i,t-1}$ and then may be subject to change multiple times, if this agent is chosen randomly multiple times as a conversation partner by other agents.

## 2.2 Advertisement

In each period, agents of type A use all of their funds (which are replenished as described in subsection 2.5) to advertise their views to $n_t^{adv}$ other agents. Since, in this simple model, money can only be spent on advertisements, we can choose the costs of sending one advertisement to one agent as the *numeraire*. Accordingly, the number of advertisements, i.e. $n_{i,t}^{adv}$, is the highest integer which is lower than or equal to the agent's funds $f_{i,t}$ (i.e. $f_{i,t}$ is rounded down to the next integer). Agents j who receive advertisements also update their beliefs according to equation (1), where $b_{i,t}^*$ is the belief of the advertiser.

## 2.3 Information from the outside world

After agents exchanged their opinions and consumed advertisements, a percentage of each agent type $\gamma_A$ and $\gamma_B$ receives perfect information about their true interests. In the context of climate change, we could imagine that e.g. natural disasters and extreme heat during summers can make a certain fraction of people aware about the negative consequences of greenhouse gas emissions. Polluters, on the other hand, will feel the consequences of curbing their emissions in their bank accounts.

The final belief held in this period, $b_{i,t}$, is thus determined according to the following equation, where $i_X$ denotes true interest of an agent of type X (which may be either A or B):

$$b_{i,t} = \begin{cases} i_X & \text{for } \frac{\gamma_X}{100} n_X \text{ agents of type X } \in \{A, B\} \\ b_{i,t}^* & \text{for other agents of type X} \end{cases} \qquad (2)$$

In order to make the analysis as simple and clear as possible, I assume polarized interests, i.e. all non-polluters have the largest payoff for a law $l_t$ of 0. This means that their true interests are $i_B = 0$, whereas all agents of type A receive maximum profits at $l_t = 1$, i.e. their true interests are $i_A = 1$. Please note that this specification does *not* exclude cases in which the technical minimum or maximum lies between 0 and 1. As long as either $\gamma_A > 0$ or $\gamma_B > 0$ holds, i.e. at least some agents have a positive probability of receiving perfect information about the outside world, only beliefs which are located between the two interests (i.e. $i_A \geq b_{i,t} \geq i_B$) are able to prevail in the long run. Hence, we can normalize the policy space such that the true interests are located at 0 and 1 without loss of generality.

*2.4 Political decision-making*

I assume that the law $l_t$ is determined according to the Condorcet rule. This means that the law is determined by comparing each potential value of $l_t$ against each other possibility. The one value for $l_t$ that wins a majority against all other alternatives is chosen as a so-called Condorcet winner. If preferences are single-peaked and all voters participate in the election, the alternative favored by the median voter wins against all other alternatives (Black 1948). Hence, the median belief $b_t^m$ determines policy:

$$l_t = b_t^m \qquad (3)$$

*2.5 Profits*

For simplicity, the profits of agents of type A $\pi_t$ are determined by a linear function of the policy $l_t$ and the maximum profits $\pi^{max}$:

$$\pi_t(l_t) = l_t \pi^{max} \qquad (4)$$

Profits replenish the funds of agents of type A, which are used to buy advertisements (see subsection 2.2). The funds that each agent of type A is able to command are thus updated according to the following equation:

$$f_{i,t+1} = f_{i,t} - n_{i,t}^{adv} + \pi_t \tag{5}$$

*2.6 Costs*

It is technically not necessary to make assumptions about the costs that agents of type B have to bear for a given law within our model. Doing so, however, helps to a) rationalize the model setup and b) interpret the model outputs in terms of efficiency and equity.

I assume that the costs are represented by a function $g(l_t)$ and that total costs are more than double the total profits made by agents of type B for any policy $l_t > 0$, i.e.:

$$n_B g(l_t) > 2 n_A \pi_t(l_t) \tag{6}$$

I further assume that $g(l_t)'' > 0$, i.e. the marginal costs imposed on agents of type B are increasing. Again, this assumption seems to be reasonable in the context of climate change, as each additional degree of global warming is expected to trigger more severe consequences.

This specification implies that agents of type A cannot "buy" a majority of agents of type B under full rationality and perfect information. It also implies that any level of $l_t > 0$ is both unequal and inefficient. Furthermore, the inefficiency increases with $l_t$, i.e. an $l_t = 1$ (determined by the median belief) is both most unequal and most inefficient.

# 3 Results

As it is standard in this stream of literature, I study the properties of my model with the help of numerical simulations. Since both the setup of the network of agents and the interactions between the agents are based on stochastic processes, I conduct several simulations per parameter combination in order to ensure that my analysis does not rely on outliers. I furthermore perform extensive sensitivity analysis in order to explore how the model's parameters influence the results.

The baseline calibration is described in table 1. In order to emphasize the effects that even a tiny minority may have, the number of type A agents is set to one.

**Table 1: Parameter setting**

| Parameter | Symbol | Value |
|---|---|---|
| Number of type A agents | $n_A$ | 1 |
| Number of type B agents | $n_B$ | 999 |
| True interests of A | $i_A$ | 1 |
| True interests of B | $i_B$ | 0 |
| Initial belief (mean) | - | 0.5 |
| Initial belief (standard deviation) | - | 0 |
| Share of type A agents who receive perfect information | $\gamma_A$ | 100% |
| Persuasion parameter | $\mu$ | variable (0.025-1) |
| Share of type B agents who receive perfect information | $\gamma_B$ | variable (0.025%-15%) |
| Connection probability (random network) | - | 2.5% |
| Neighborhood size (Watts-Strogatz network) | - | 5 |
| Rewire probability (Watts-Strogatz network) | - | 10% |
| Maximum profits | $\pi^{max}$ | Variable (0-500) |

### 3.1 Perfect information

If both sides are perfectly informed about their situation, i.e. $\gamma_A = 1$ and $\gamma_B = 1$ type B agents will adhere to the belief 0 and type A agents to 1. Since the majority of agents belongs to type B, the median belief and accordingly also policy $l_t$ will be 0. This outcome is both efficient and equal.

### 3.2 One side is perfectly informed, the other not at all

If $\gamma_A = 1$, but $\gamma_B = 0$, the belief 1 will spread until the whole population adheres to it, if there exists some connection between any agents x and y (over an arbitrary number of intermediate agents). This process follows a concave curve whose exact shape depends not only on the persuasion parameter, but also on the network structure, as a low connectedness between the agents may slow down the adoption of this belief (see fig. 1).

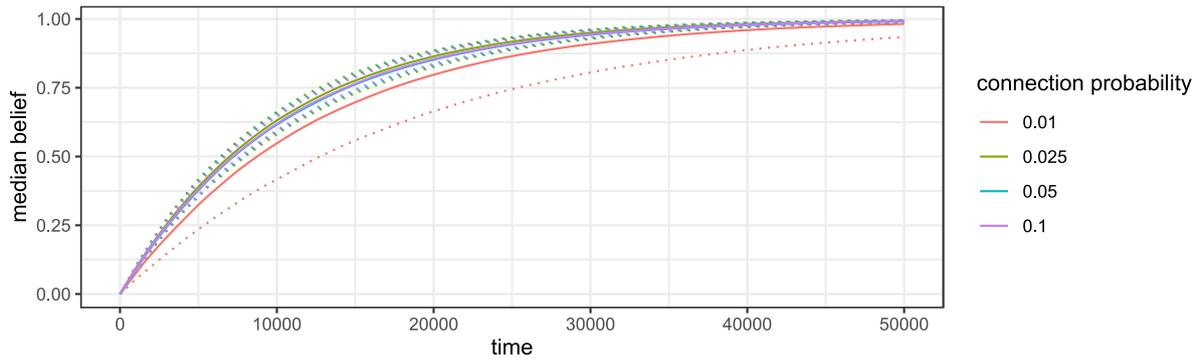

**Figure 1:** The evolution of median belief for different connection probabilities in a random network. Quantile regressions of 50 simulations per parameter combination.

### 3.3 One side is better informed

What if some agents of type B also receive information about their true interests, i.e. $\gamma_B > 0$? Figure 2 shows that the median belief (after 50,000 simulation periods) can be larger than 0 if $\gamma_B$ is low. This means that suboptimal outcomes can persist if information inequality, i.e. inequality in the ability to obtain information about one's true interests, is high. This relationship is affected by the persuasion parameter $\mu$, as an increase in $\mu$ increases the median belief for low $\gamma_B > 0$. However, the median belief is still close to its optimal value of 0 for these simulations.

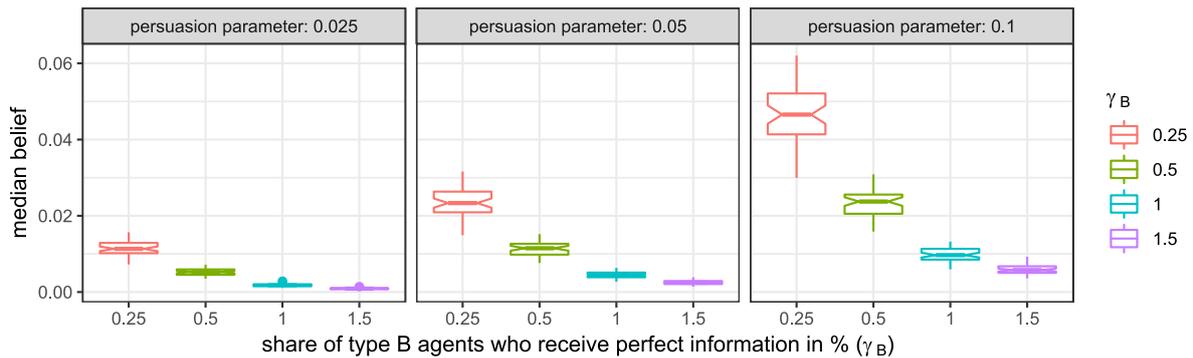

**Figure 2:** Median belief after 50,000 periods for different levels of information inequality and persuasion parameters. Notched boxplots of 50 simulations per parameter combination.

*3.4 Advertisements*

I finally allow the agents of type A to advertise their views using their profits under varying parameter values for maximum profit, and compare three different advertisement strategies:

a) *random targeting*, where each agent has an equal probability of receiving advertisements,

b) *efficient targeting*, where those agents who adhere the least to $i_A$, i.e. where the advertisement has the largest impact on the agent's belief, receive the advertisements, and

c) *influencer targeting*, where the advertisements are sent to those agents who have the most connections to other agents.

Figure 3 shows the evolution of the median belief for different initial beliefs. Although path-dependency may play a role for a while, the median belief ultimately settles at a steady state which converges to a quasi-steady state in the long-run for any simulation run with a given persuasion parameter (in this case: 0.025), maximum profit (here: 500), share of agents of type B who receive a signal about their true preferences (here: 0.5%) and advertisement strategy.[3]

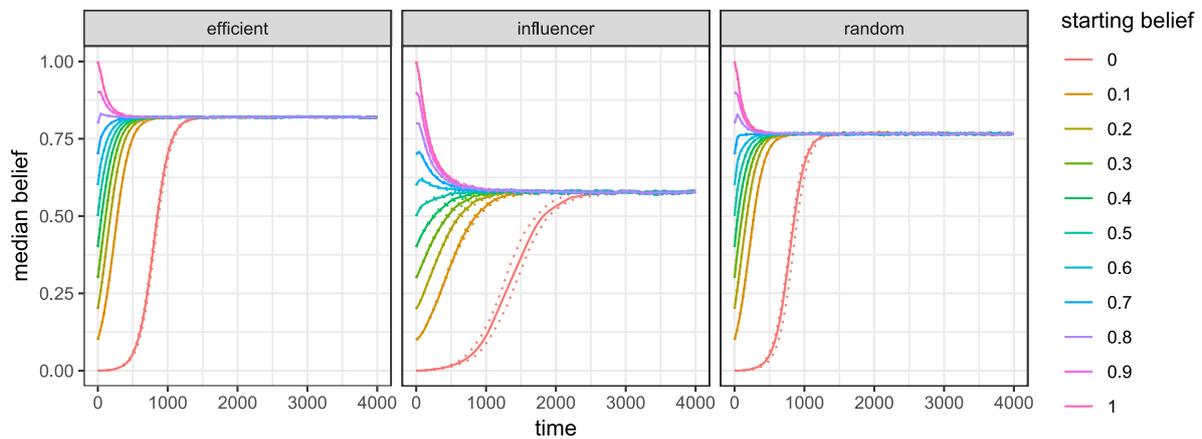

**Figure 3:** The evolution of median belief for different initial beliefs. Quantile regressions of 10 simulations per parameter combination.

---

[3] Some fringe cases with (unrealistically) high values for the persuasion parameter exist where public opinion does *not* settle to an equilibrium, see fig. 4.

Figure 4 shows the density of beliefs for a single simulation run per advertisement strategy. The "influencer" targeting exhibits three stable poles: a) around "influencers", who consume advertisements and are thus relatively close to belief of type A, b) those who have recently received a signal about their true preferences are located around belief 0, and c) other agents who are not consuming advertisements, but are influenced by the two poles. If the agent of type A pursues a "random" or "efficient" advertisement strategy, on the other hand, agents are confined to two poles: around the median voter and around 0.

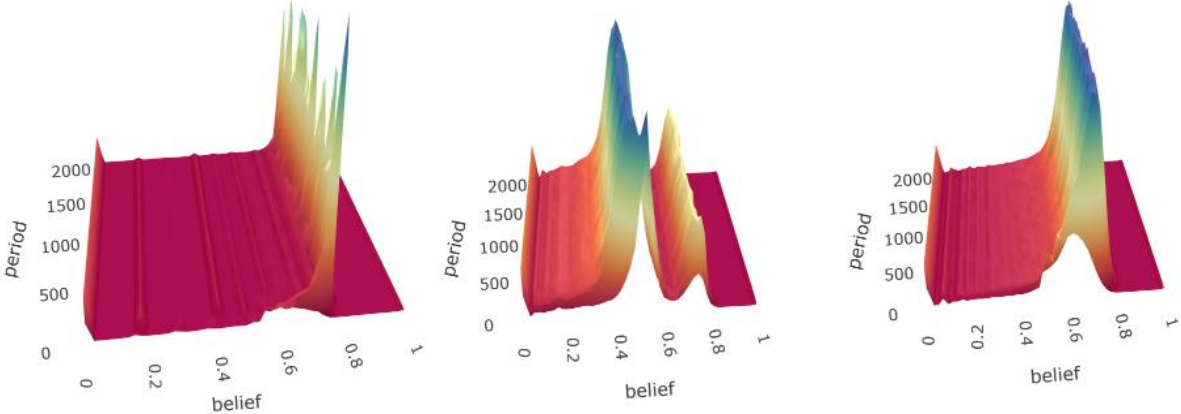

**Figure 4:** Density of beliefs for a single simulation run per advertisement strategy according a kernel density estimation. Left: "efficient", center: "influencer", right: "random"

Figure 5 shows the median belief (i.e. the policy) held on average between period 4501 and 5000 as mean of ten simulation runs for a "random" advertisement strategy. It can be seen that an increase in maximum profits drastically increases the parameter space that allows for a median belief which is larger than 0. Furthermore, there is a non-linear relationship between the persuasion parameter and the median belief: a persuasion parameter of 0.5 allows the agent of type A to have the largest influence on public opinion according to this agent's interests, if we assume that the other parameters stay constant.

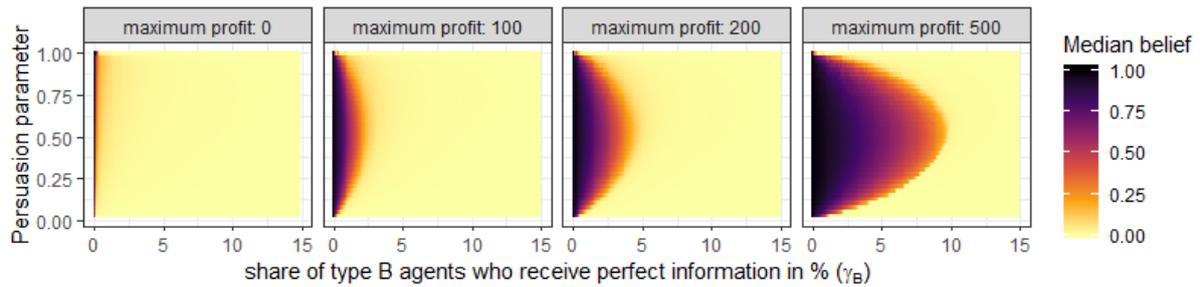

**Figure 5**: Median belief, i.e. law, held on average between period 4501 and 5000 for "*random*" advertisements as mean of 10 simulation runs per parameter combination for a random network

This counterintuitive result stems from the fact that the agent of type A can temporarily be persuaded to hold – and advertise – a belief that differs from 1. Since this agent receives a signal about his or her true preferences in every period and each agent is on average the target of one persuasion attempt, the agent of type A will on average only advertise a belief which is lower than 0.5, if the persuasion parameter is larger than 0.5.

Figure 6 illustrates the fact that opinion formation can become increasingly unstable, if the persuasion parameter is very high (in this case: 0.99). While instability may occur under certain parameter combinations, there is not too much reason to be worried, as a) figures 5, 7, 8 and 9 show that this only affects outcomes in fringe cases, and b) empirical persuasion parameters are surely much lower, as a single conversation cannot be assumed to have an impact which is that drastic.

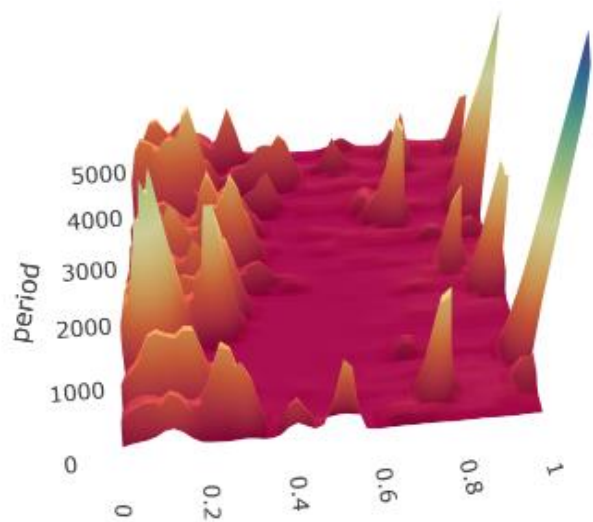

**Figure 6:** Density of beliefs for a single simulation run of a "random" advertisement strategy according to a kernel density estimation.

Figures 7 and 8 show the median belief held on average for "influencer" and "efficient" advertisements respectively. Fig. 7 shows that the "influencer" strategy has a slight comparative advantage for high levels of the persuasion parameter.

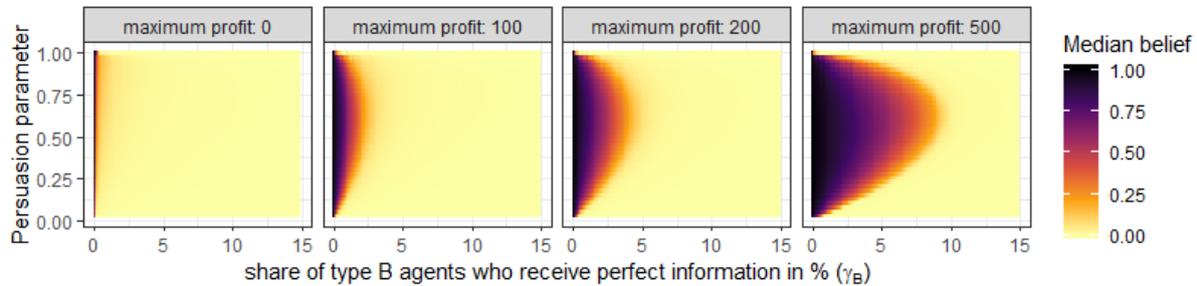

**Figure 7**: Median belief, i.e. law, held on average between period 4501 and 5000 for "*influencer*" advertisements as mean of 10 simulation runs per parameter combination for a random network

Fig. 8 shows that the "efficient" advertisement strategy enables agent the agent of type A to shift public opinion much better to their favor. It also shows that the non-linear relationship between the persuasion parameter and the median belief ceases to exist for intermediate levels of maximum profits. The driving force behind this result is that the "efficient" strategy always targets agents with beliefs closer to 0 first. Thus, even if the agent of type A is "turned" and advertises views which are against her interests, she will "preach to the converted" first, thus posing no real damage to their true interests. If maximum profits are very high, however, the advertisements will also reach out to agents which hold a belief close to 1, thus undermining the true interests. Again, these fringe cases seem only to be of a theoretical interest, as the empirical persuasion parameter can be assumed to be much smaller than 0.5 and the agent of type A will thus neither hold, nor advertise views which are far away from her actual interests.

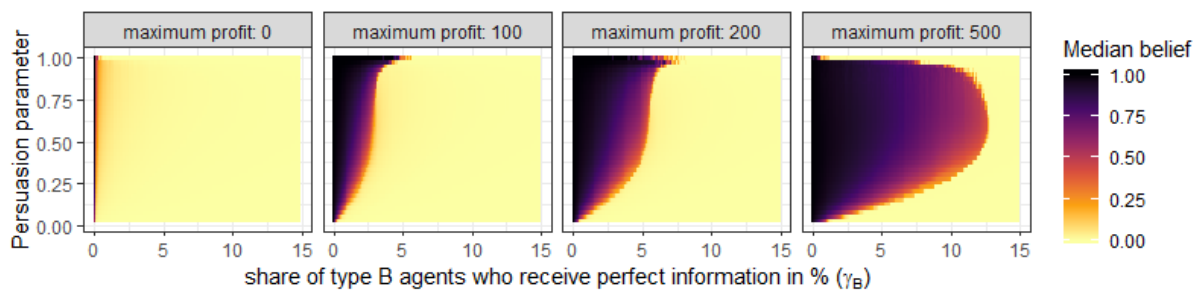

**Figure 8**: Median belief, i.e. law, held on average between period 4501 and 5000 for "*efficient*" advertisements as mean of 10 simulation runs per parameter combination for a random network

As a robustness check, I simulate a Watts-Strogatz small world network (Watts and Strogatz 1998) instead of a random network and show the results for an "efficient" advertisement strategy and show the results in fig. 9. Changing the network structure does not drastically change the results, although the parameter space that allows for inefficient outcomes becomes slightly larger in the Watts-Strogatz network (see fig. 9)

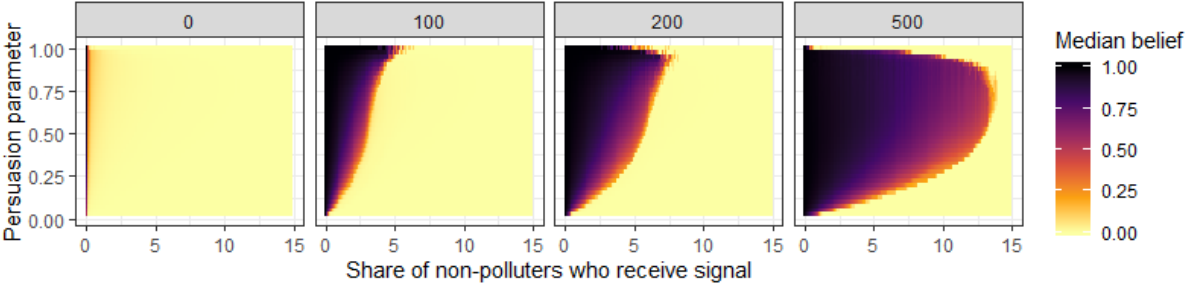

**Figure 9**: Median belief, i.e. law, held on average between period 4501 and 5000 for "*efficient*" advertisements as mean of 10 simulation runs per parameter combination for a Watts-Strogatz network.

Finally, I explore the impact of information and economic inequality on opinion polarization measured as the standard deviation of beliefs held in the population. In the absence of advertisements, polarization is generally low irrespective of the median belief.[4] If agents of type A can invest their funds into advertisement, however, this does not only have an increasing effect on the median belief, but also on the dispersion of beliefs (see fig. 10). Thus, the model suggests a channel from economic inequality to both inefficient policy and political polarization.

---

[4] If the persuasion parameter is (close to) 1, opinions may be polarized, as the beliefs held in the population will then be (close to) 1 or (close to) 0, mechanically pushing the standard deviation of beliefs upwards.

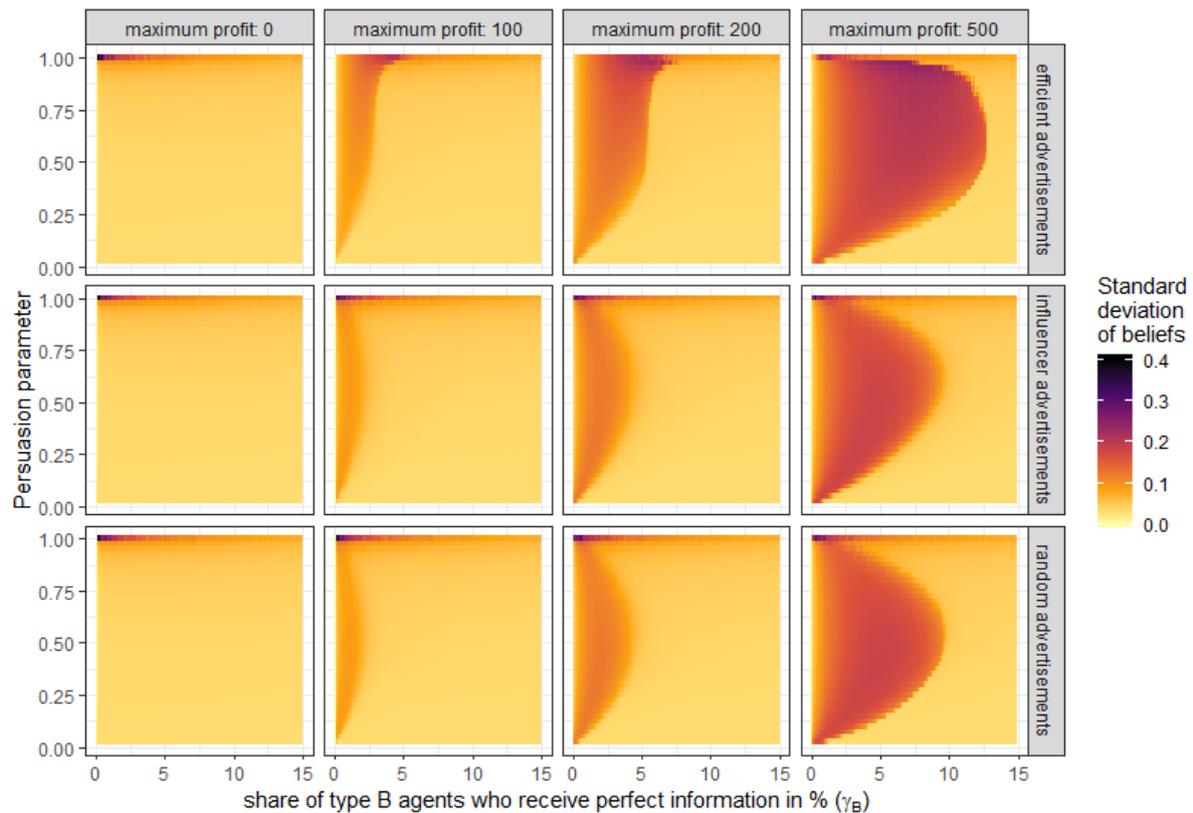

**Figure 10**: Standard deviation of beliefs held on average between period 4501 and 5000 as mean of 10 simulation runs per parameter combination for a random network.

# 4 Conclusion

I have developed a parsimonious agent-based model that allows to capture an interplay between opinion formation, political decision making and the economy. The model is populated by two types of agents with conflicting interests, who may differ with regard to their ability to a) gain outside information about their true interests ("information inequality") and b) advertise their views ("economic inequality").

Using numerical simulations, I showed that information inequality can enable even a tiny minority to shift public opinion to its advantage. This effect is greatly amplified, if this group is able to afford large-scale propaganda to advertise their views and effectively disinform the majority of the population. In particular, "targeted advertisement" strategies as made possible by the rise of information technology and, especially, social media, are particularly effective in spreading misinformation and fostering political polarization.

Within the model, shifting public opinion triggers changes in the law, which in turn provides the minority with more economic resources that can be used for disinformation campaigns. Hence, the model illustrates how information inequality can produce self-reinforcing economic inequality within a democracy. My model further suggests that economic inequality can be a driver of inefficiency and political polarization, as disinformation campaigns can cause a large share of the population to adopt misinformed beliefs.

My model aims, among others, to capture some of the salient features of the opinion formation about climate change. Recent research suggests that major players in the oil industry were aware of the causes, extent, possible trajectories and damages of climate change well before the majority of the population. They used this advantage in information as well as their profits to finance disinformation campaigns (e.g. Frumhoff et al. 2015; Bonneuil et al. 2021), hence providing a showcase for my model.

My model suggests two ways to simultaneously combat a) mis- and disinformation, b) political polarization, c) economic inefficiency and d) economic inequality: First, by reducing information inequality by providing a larger share of the population with access to unbiased information, e.g. by providing funding to i) media outlets, and ii) science that is independent from company profits and government agenda. Second, by reducing economic inequality or its impact, e.g. by i) increasing wealth-related taxes and/or ii) limiting the extent of advertisement/propaganda in general and disinformation in particular.

The model presented in this paper only provides a simplified account of opinion formation, collective decision-making and the economy, and can thus be extended in various ways. First, it would be interesting to study the effects of a more complex specification of opinion dynamics. For instance, a "bounded confidence interval" (Hegselmann and Krause 2002) could provide some defense against disinformation (see also Douven and Hegselmann 2021). On the other hand, disinformation campaigns could use "salami tactics" (Schelling 1966) to circumvent this defense by moderating the views advertised to an extent that is able to invade the bounded confidence interval of the median voter. Modelling the economy in more detail seems to be another promising avenue for further research, as this could allow for a more complex co-evolution of opinion formation and economic outcomes. Furthermore, one could introduce more complexity in the political sector by introducing, for instance, politicians who pursue policies which accord to their own preferences even though they are not backed by the majority. Such a mechanism could have manifold interpretations, by representing either self-sacrificing politicians who aim to provide merit goods, or ideological fanatics, or corrupt politicians who are bought off by some special interest groups. Finally, in the context of climate change, the share of agents who receive a signal about their true interests could be endogenized by adding an environmental module to the model. The signal could then depend on the status of environmental degradation, which itself could depend on the law chosen in the previous periods. One could then use this model to study opinion tipping points and under which conditions they occur "too late" to save the environment.

**Competing interests**: The author declares that he has no competing interests.

**Acknowledgements**: I thank Christian Gehrke for valuable comments on a previous version of this manuscript and Dagmar Holzschuster for proofreading it. All errors are mine.